# High thermoelectric performance in the hexagonal bilayer structure consisting of light boron and phosphorus elements

Z. Z. Zhou, H. J. Liu[*], D. D. Fan, G. H. Cao, C. Y. Sheng

*Key Laboratory of Artificial Micro- and Nano-Structures of Ministry of Education and School of Physics and Technology, Wuhan University, Wuhan 430072, China*

Two-dimensional layered materials have attracted tremendous attentions due to their extraordinary physical and chemical properties. Using first-principles calculations and Boltzmann transport theory, we give an accurate prediction of the thermoelectric properties of boron phosphide (BP) bilayer, where the carrier relaxation time is treated within the framework of electron-phonon coupling. It is found that the lattice thermal conductivity of BP bilayer is much lower than that of its monolayer structure, which can be attributed to the presence of van der Waals interactions. On the other hand, the graphene-like BP bilayer shows very high carrier mobility with a moderate band gap of 0.88 eV. As a consequence, a maximum *p*-type *ZT* value of ~1.8 can be realized along the *x*-direction at 1200 K, which is amazingly high for systems consisting of light elements only. Moreover, we obtain almost identical *p*- and *n*-type *ZT* of ~1.6 along the *y*-direction, which is very desirable for fabrication of thermoelectric modules with comparative efficiencies. Collectively, these findings demonstrate great advantages of the layered structures containing earth-abundant elements for environment-friendly thermoelectric applications.

## 1. Introduction

In the past decades, two-dimensional (2D) materials have attracted tremendous attentions due to their intriguing properties and great potential in various applications such as deep ultraviolet laser, optoelectronic device, and energy conversion [1−6]. In order to develop high performance thermoelectric materials which can directly convert waste heat into electrical power, 2D systems require moderate band gap, reasonably large carrier mobility, and relatively low thermal conductivity. Unfortunately, the thermoelectric applications of most 2D materials are more or less limited by some inherent weaknesses. For example, graphene has been investigated intensively for its ultrahigh carrier mobility and massless Dirac fermions, while the gapless energy band makes it unsuitable for the application as thermoelectric material [7]. In contrast, the graphene-like III–V binary compounds are all predicted to exhibit direct band gaps [8], whereas the representative hexagonal boron nitride (BN) monolayer, which has been successfully isolated, behaves as an insulator with a wide band gap of about 6.0 eV [9−13].

---

[*] Author to whom correspondence should be addressed. Electronic mail: phlhj@whu.edu.cn

Recently, another III–V binary compound named hexagonal BP monolayer has also attracted considerable attentions [14−18]. Experimentally, the BP films with wurtzite structure have been synthesized by chemical vapor deposition [19]. Similar to BN monolayer, it is believed that BP layers could also adopt hexagonal lattice if prepared appropriately [20]. Theoretically [21], the phonon dispersion relations of BP monolayer show no imaginary frequency, and its in-plane Young modulus and Poisson's ratio are comparable with those of the $MoS_2$ monolayer. All these ensure a satisfied mechanical stability of BP monolayer. Besides, Wang *et al.* [22] found that the B−P bond is still reserved in the Born-Oppenheimer molecular dynamics simulation for 5 ps at the temperature of 2500 K, which suggests high thermal stability of BP monolayer. On the other hand, it was reported that BP monolayer is a semiconductor with low effective mass and moderate band gap [23,24]. It is thus naturally reminiscent of the thermoelectric application of BP monolayer, whereas an inherent problem is that almost all of the hexagonal monolayer structures consisting of light elements have pretty large lattice thermal conductivities [25,26]. Fortunately, previous investigations revealed that the lattice thermal conductivities of bilayer graphene and hexagonal BN are obviously lower than those of their monolayer counterparts [27−31]. Accordingly, we believe that BP bilayer should also have lower lattice thermal conductivity and thus better thermoelectric performance compared with BP monolayer, and a comprehensive understanding on its electronic and phonon transport properties is quite necessary.

In this work, the thermoelectric transport properties of the hexagonal BP bilayer are investigated by using first-principles calculations and Boltzmann theory. It is demonstrated that the lattice thermal conductivity of BP bilayer is much lower than that of monolayer structure due to the van der Waals (vdW) interactions. Combined with the extremely large power factors originated from the ultrahigh carrier mobility, an unexpected high thermoelectric performance can be realized in both *p*- and *n*-type BP bilayer in the high temperature region.

## 2. Computational method

The lattice thermal conductivity ($\kappa_l$) of the BP bilayer can be obtained by solving the phonon Boltzmann transport equation, as implemented in the so-called ShengBTE package [32]. The second- and third-order interatomic force constants (IFCs) are investigated by density functional theory (DFT) combined with the finite displacement method using a 5×5×1 supercell. The ninth nearest neighbors are included for the third-order interactions to ensure convergence. The DFT calculations are implemented in the Vienna *ab-initio* simulation package (VASP) [33], and the harmonic and anharmonic IFCs are computed based on the PHONOPY program [34] and the THIRDORDER.PY script [32], respectively.

The electronic properties of the BP bilayer are calculated within the framework of DFT, as implemented in the QUANTUM ESPRESSO package [35]. The norm-conserving scalar-relativistic pseudopotential is used to describe the core-valence interaction [36], and the exchange-correlation functional is in the form of Perdew-Burke-Ernzerhof (PBE) with the generalized gradient approximation (GGA) [37]. To eliminate the interactions between the bilayer and its periodic images, the system is modeled by adopting hexagonal supercell with the vacuum distance of 30 Å. The vdW interactions are treated by including the optB86b functional when optimizing the lattice parameters [38]. The kinetic energy cutoff is set to 80 Ry and the atomic positions are fully relaxed. To give accurate electronic transport coefficients, the band structure of BP bilayer is obtained by adopting hybrid density functional in the form of Heyd-Scuseria-Ernzerhof (HSE) [39], which has been successfully used to predict the band gaps of bulk BP [22] and hexagonal BN monolayer [40]. Based on the calculated band structure, the Seebeck coefficient ($S$), the electrical conductivity ($\sigma$), and the electronic thermal conductivity ($\kappa_e$) can be derived from the Boltzmann transport theory [41], where the key point is appropriate treatment of the carrier relaxation time $\tau$. Using the density functional perturbation theory (DFPT) [42] and the Wannier interpolation techniques [43], the $k$ resolved relaxation time can be obtained from the imaginary part of the electron self-energy by a complete electron-phonon coupling (EPC) calculation [44]. To ensure converged results, the EPC calculations have been performed by using coarse grids of $20\times20\times1$ $k$-points with $10\times10\times1$ $q$-points, and then interpolated to dense meshes of $200\times200\times1$ $k$-points with $100\times100\times1$ $q$-points via the maximally localized Wannier functions, as implemented in the electron-phonon Wannier (EPW) package [44]. To make a better comparison, the transport coefficients $\sigma$, $\kappa_e$, and $\kappa_l$ are all renormalized with respect to the interlayer distance of BP bilayer.

## 3. Results and discussion

Owing to the low cleavage energy, the BP bilayers have been proved to be easily realized in experiment [22]. Among several possible bilayer systems, the stacking configuration shown in Figure 1 is the most stable one [22,45]. As can be seen from Fig. 1(a), the bottom B1 atoms entirely overlap with the top B2 atoms in the *xy* plane, while the P1 atoms are positioned at the center of the hexangular ring consisting of B2 and P2 atoms. The optimized interlayer distance is 3.57 Å, and the atoms in each layer are almost completely located in the same plane. The bilayer structure keeps the same B−P bond length (1.86 Å) as that of the BP monolayer, which is in good agreement with previous theoretical results [45].

Figure 2(a) displays the phonon spectrum of the BP bilayer, where there is no imaginary frequency guaranteeing the dynamic stability of the system. As the primitive cell of the bilayer system contains four atoms, there exist 12 phonon branches with three acoustic (LA, TA, ZA) and nine optical ones (LO$^i$, TO$^i$, ZO$^i$, $i$ = 1, 2, 3). It can be seen that the LA and TA modes show linear dependence on the wave vector near the Brillouin zone, while the ZA mode exhibits a quadratic dispersion, which are very similar to those found in layered systems such as graphene [25] and hexagonal BN monolayer [31]. As a comparison, Fig. 2(b) also plots the phonon dispersion relations of BP monolayer. If two independent monolayers are held together, one would expect a complete overlap of their phonon dispersions. Indeed, we see that Fig. 2(a) and Fig. 2(b) are very similar to each other and the doubly degenerate bands (TA and TO$^1$, LA and LO$^1$, TO$^2$ and TO$^3$, LO$^2$ and LO$^3$) remain unchanged. The only exception is that the ZA and ZO$^1$ modes exhibit obvious band splitting around the $\Gamma$ point, which is less pronounced for the ZO$^2$ and ZO$^3$ modes. Such observation suggests that the interlayer vdW interactions in the BP bilayer mainly affect the out-of-plane phonon modes. Meanwhile, we see that all the ZO modes hybridize with the acoustic modes in the BP bilayer, which means more three-phonon process and thus a lower relaxation time [46]. We will come back to this point later. Fig. 2(c) plots the lattice thermal conductivity of the BP bilayer as a function of temperature from 300 K to 1300 K. For comparison, the result of the BP monolayer is also shown in the inset. In the whole temperature range, we see that the $\kappa_l$ of the BP bilayer exhibit less isotropy and the values are nearly one order of magnitude lower than those of the monolayer structure. Table 1 summarizes the contribution of different phonon modes to the lattice thermal conductivity at a prototypical temperature of 1200 K. When the system changes from the monolayer to bilayer, we see there is a significant reduction of the lattice thermal conductivity contributed by the acoustic modes, especially for the TA and ZA phonons. In another word, the heat transport by the acoustic phonons is largely limited in the BP bilayer. To have a further understanding, we first note that the phonon group velocities of BP bilayer and monolayer are almost identical to each other (see Figure S1(a) and S1(b) in the supplementary materials). This is reasonable since the phonon dispersion relations in both systems are very similar. We further observe from Fig. 2(d) and its inset that the relaxation times of the acoustic phonons in the BP bilayer are at least one order of magnitude lower than those in the BP monolayer, which can be attributed to the fact that more optical branches hybridizes with acoustic branches (see Fig. 2(a)) caused by the vdW interactions discussed above. Besides, the relaxation times of the low frequency optical phonons are also obviously lower for the bilayer system. As a consequence, the remarkably lower lattice thermal conductivity of the BP bilayer can be attributed to its significantly lower phonon relaxation time caused by the presence of vdW interactions. It should be noted that the EPC may also have certain influences on the phonon transport properties [47−50], especially at high carrier concentration. However, our additional calculations (see Figure S2 in the supplementary materials) find that the phonon relaxation time from the intrinsic phonon-phonon scattering is at least two orders of magnitude lower than that originating from the EPC. It is thus reasonable to neglect the effects of EPC on the lattice thermal conductivity of the BP bilayer.

Figure 3(a) displays the energy band structure of the BP bilayer calculated by adopting the HSE functional. In contrast to that of the BP monolayer (see Figure S3 in the supplementary materials), the conduction band minimum (CBM) of the bilayer system is located at the **K** point whereas the valence band maximum (VBM) is slightly shifted away. The HSE bands show an indirect gap of 0.88 eV, which is more suitable for thermoelectric application compared with that of the monolayer system (1.37 eV). Besides, we see there are two top valence bands degenerated at the **K** point, which indicates a sharp increase in the density of state (DOS) near the VBM and thus a large *p*-type Seebeck coefficient is expected [51]. In contrast, the steep energy dispersion around the CBM means a relatively lower *n*-type Seebeck coefficient but a higher group velocity. Note that the dispersions around the VBM and CBM of BP bilayer are similar to those of the monolayer, suggesting that the carrier mobility of BP bilayer should be comparable with that of the monolayer structure which has been proved to be very high in previous work [24].

We now move to the discussion of the carrier relaxation time of the BP bilayer. It should be noted that better thermoelectric performance should appear at higher temperatures due to its relatively larger band gap compared with those of the $Bi_2Te_3$ [52] and SnSe [53]. In Fig. 3(b), we plot the energy dependence of the carrier relaxation time at 1200 K. It is found that the BP bilayer exhibits higher *n*-type carrier relaxation time, which can be attributed to less scattering channels near the CBM originating from the lower DOS. The relaxation times near the VBM and CBM are calculated to be 21 fs and 42 fs at 1200 K, respectively, which are obviously higher than those of $Bi_2Te_3$ [52] and suggest very favorable electronic transport properties of the bilayer system. Based on the calculated band structure and the carrier relaxation time, the electronic transport coefficient of the BP bilayer can be accurately predicted by solving the Boltzmann transport equation. Figure 4(a) and 4(b) show the absolute values of the Seebeck coefficient and the electrical conductivity of BP bilayer at 1200 K, respectively. In the concentration range of $3\times10^{12} \sim 3\times10^{13}$ cm$^{-2}$, we find obvious higher value of *p*-type Seebeck coefficient caused by the band degeneracy around the VBM while larger *n*-type electrical conductivity due to the strong band dispersion near the CBM. As a consequence, we see from Fig. 4(c) that the BP bilayer shows similar *p*- and *n*-type power factors ($S^2\sigma$) at 1200 K, especially for those along the *y*-direction. At the optimized electron concentration of $4\times10^{12}$ cm$^{-2}$, the *y*-direction Seebeck coefficient of −228 μV/K for the BP bilayer is comparable with those found in many good thermoelectric materials [52,53], where the electrical conductivity of the former is much larger (3113 S/cm) due to its ultrahigh carrier mobility (6294 cm$^2$/Vs). Extraordinarily high power factor ($1.6\times10^{-2}$ W/mK$^2$) can thus be realized at 1200 K, suggesting the great advantage of the BP bilayer for thermoelectric application.

With all the transport coefficients obtained, we can now evaluate the thermoelectric performance of the BP bilayer, which is characterized by the dimensionless figure-of-merit $ZT = S^2\sigma T/(\kappa_e + \kappa_l)$. Fig. 4(d) displays the calculated *ZT* values as a function of carrier concentration at 1200 K, where we see that both the *p*- and *n*-type BP bilayer exhibit good thermoelectric performance. The highest *ZT* value of 1.8 can be achieved along the *x*-direction at the hole concentration of $1.2\times10^{13}$ cm$^{-2}$, which is amazingly high for systems consisting of light elements only. For the case of *y*-direction, we see that both the *p*- and *n*-type systems exhibit almost identical *ZT* value of ~1.6, which is very desirable for fabrication of thermoelectric modules with comparative efficiencies. The optimized *ZT* values of the BP bilayer are summarized in Table 2 together with the corresponding carrier concentration and transport coefficients. On the other hand, although the best thermoelectric performance appears at 1200 K, our addition calculations (see Figure S4 in the supplementary materials) show that fairly good *ZT* values above 1.0 can already be achieved at 700 K, indicating that the BP bilayer can also be operated in the intermediate temperature region. Compared with traditional thermoelectric materials, the BP bilayer, only consisting of light, earth-abundant, and environment-friendly elements, may be very favorable candidate to be designed as promising thermoelectric materials.

## 4. Summary

We present a comprehensive study of the thermoelectric properties of the BP bilayer, which shows a much lower lattice thermal conductivity compared with that of the monolayer structure caused by the presence of the vdW interactions, as well as a quite large power factor originated from the ultrahigh carrier mobility. As a consequence, excellent thermoelectric performance can be achieved in the bilayer system, where almost identical *ZT* values of *p*- and *n*-type systems are found in a wide temperature range, indicating great advantage of the fabrication of thermoelectric modules with comparative efficiencies. It is interesting to check whether other hexagonal layered structure in the III–V compounds could also exhibit good thermoelectric performance by appropriate design of the constituent elements and stacking order, which will be addressed in our future work.

**Acknowledgements**

We thank financial support from the National Natural Science Foundation (Grant Nos. 51772220 and 11574236). The numerical calculations in this work have been done on the platform in the Supercomputing Center of Wuhan University.

**Table 1** The contributions of different phonon modes to the lattice thermal conductivity (in unit of W/mK) of the BP monolayer and bilayer at 1200 K.

|  | direction | $\kappa_{ZA}$ | $\kappa_{TA}$ | $\kappa_{LA}$ | $\kappa_{OP}$ |
|---|---|---|---|---|---|
| monolayer | x | 12.1 | 21.3 | 13.8 | 8.2 |
|  | y | 30.9 | 16.7 | 14.1 | 9.4 |
| bilayer | x | 0.5 | 1.3 | 1.1 | 4.2 |
|  | y | 0.4 | 0.3 | 1.7 | 5.2 |

**Table 2** The optimized *ZT* values of the BP bilayer at 1200 K. The corresponding carrier concentration and transport coefficients are also listed.

|  | carriers | $n$ ($10^{12}$ cm$^{-3}$) | $S$ (μV/K) | $\sigma$ (S/cm) | $S^2\sigma$ ($10^{-3}$ W/mK$^2$) | $\kappa_e$ (W/mK) | $\kappa_l$ (W/mK) | ZT |
|---|---|---|---|---|---|---|---|---|
| x | hole | 12 | 237 | 3001 | 17 | 4.4 | 7.1 | 1.8 |
|  | electron | 4.1 | −228 | 2675 | 14 | 4.2 | 7.1 | 1.5 |
| y | hole | 13 | 233 | 3030 | 16 | 4.7 | 7.6 | 1.6 |
|  | electron | 4.0 | −228 | 3113 | 16 | 4.3 | 7.6 | 1.6 |

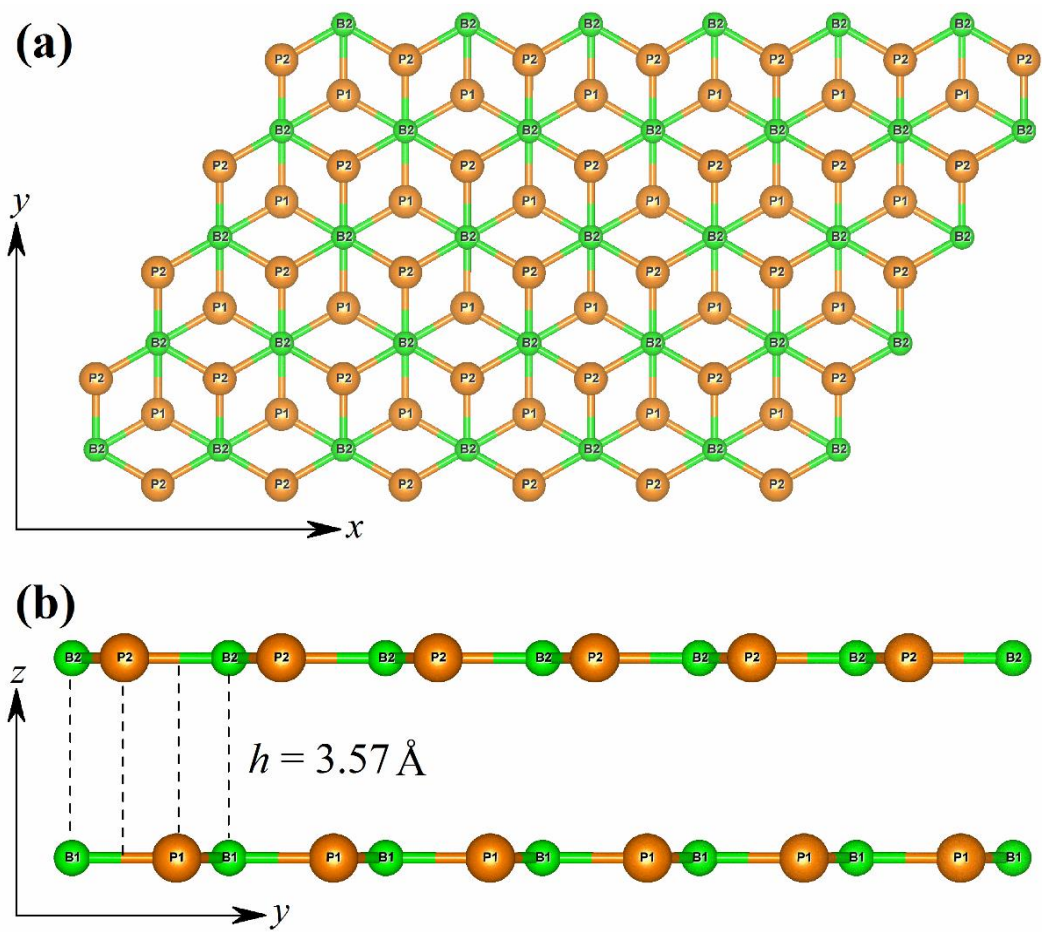

**Figure 1.** Ball-and-stick model of BP bilayer: (a) top-view, and (b) side-view.

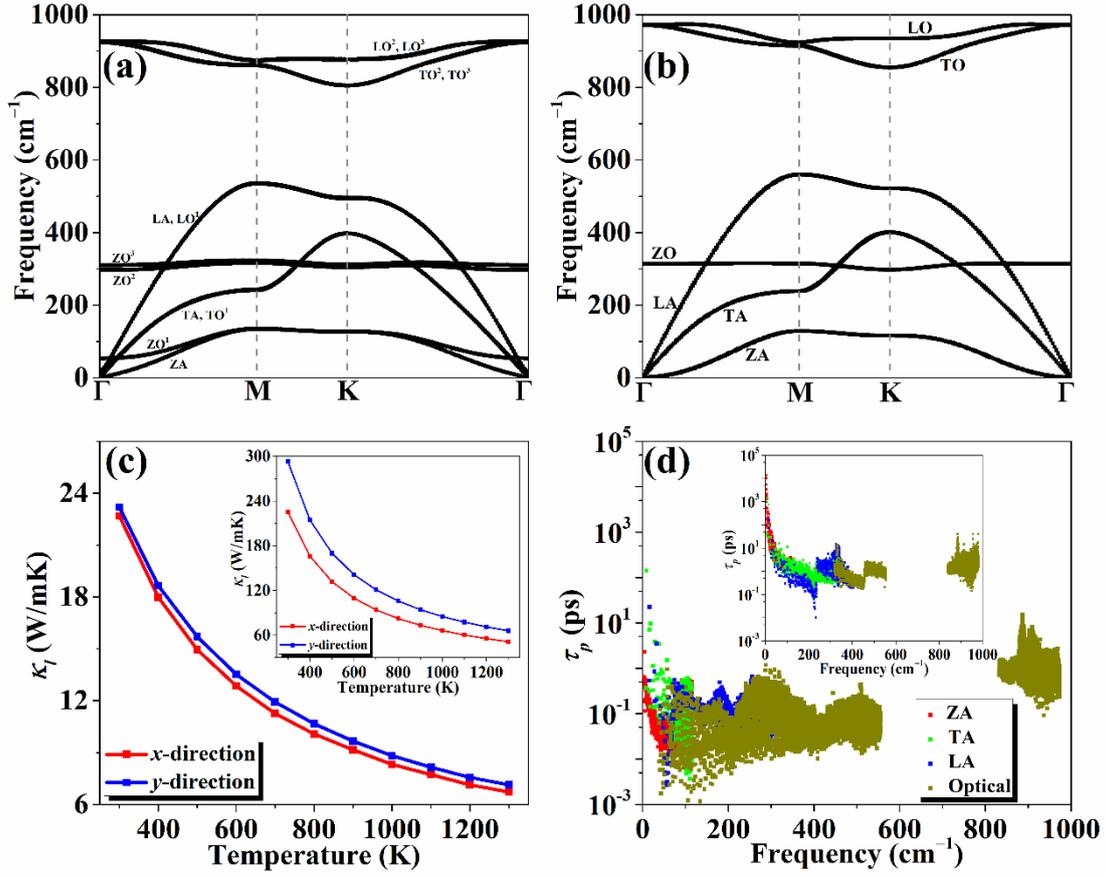

**Figure 2**. The phonon dispersion relations of the BP (a) bilayer and (b) monolayer. (c) The lattice thermal conductivity of BP bilayer as a function of temperature. (d) The phonon relaxation time of the BP bilayer as a function of frequency. The insets in (c) and (d) are the lattice thermal conductivity and the phonon relaxation time of the BP monolayer, respectively.

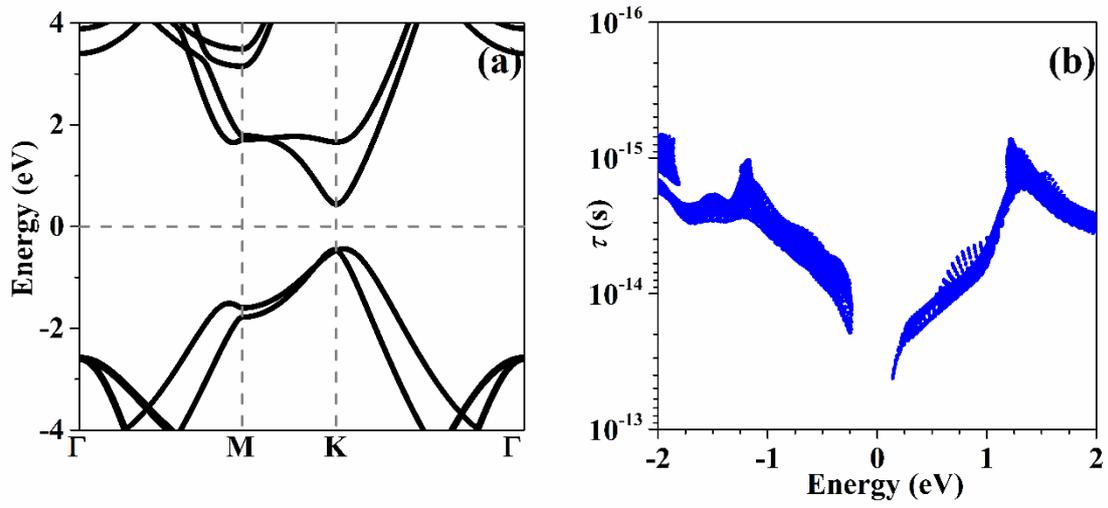

**Figure 3**. (a) The electronic band structures of BP bilayer calculated by adopting HSE functional. (b) The energy-dependent carrier relaxation time of BP bilayer at 1200 K. The Fermi level is at 0 eV.

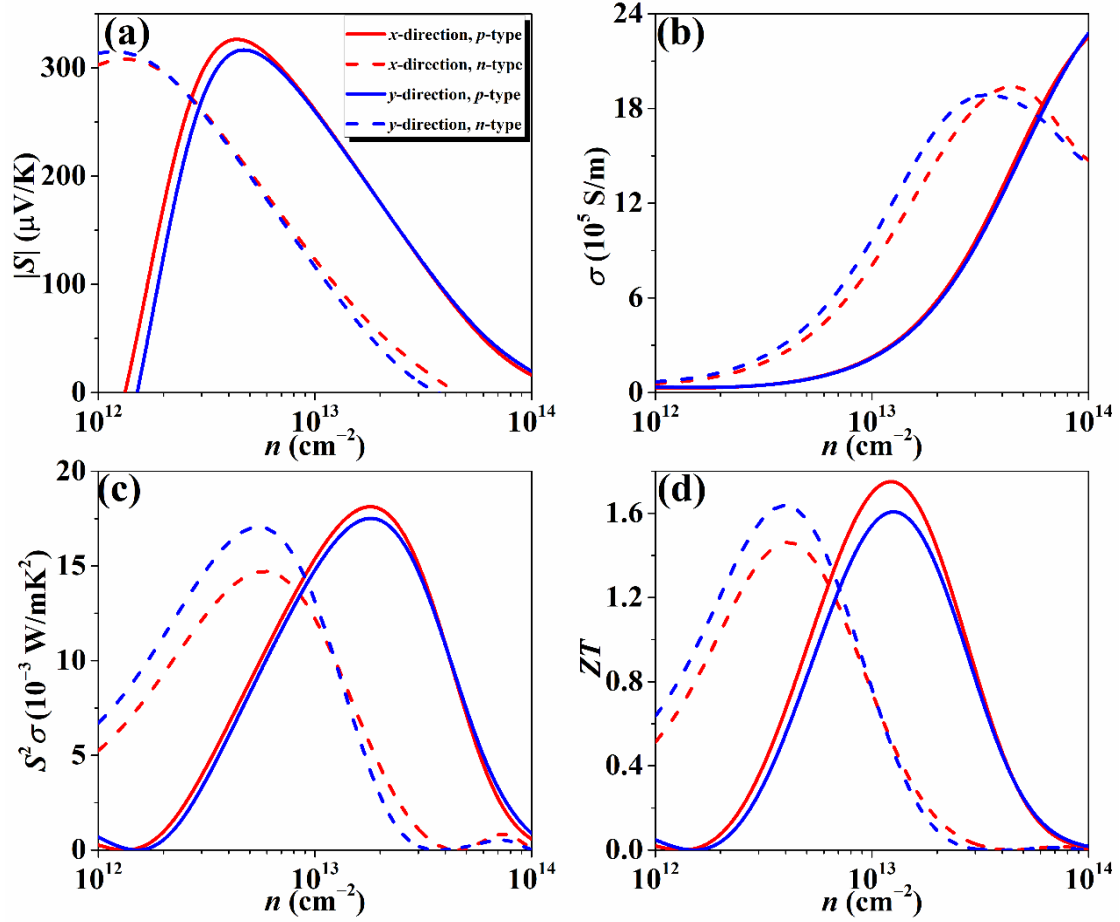

**Figure 4.** (a) The absolute values of the Seebeck coefficient, (b) the electrical conductivity, (c) the power factor, and (d) the *ZT* values of BP bilayer, plotted as a function of carrier concentration at 1200 K.